\documentclass[preprint,amsmath,amssymb]{revtex4}

\usepackage{graphicx}
\usepackage{dcolumn}
\usepackage{bm}
 
\newcommand{\mean}[1]{\langle{#1}\rangle}

\newcommand{\ket}[1]{|{#1}\rangle}

\newcommand{\dgg}{^{\dagger}}

\newcommand{\Tr}{{\rm Tr}\hspace{0.07cm}}

\begin{document}
 
 
\title{
Pure Gaussian state generation via dissipation: 
\\ A quantum stochastic differential equation approach
}

\author{Naoki Yamamoto}
\email{yamamoto@appi.keio.ac.jp}
\affiliation{%
Department of Applied Physics and Physico-Informatics, 
Keio University, 
Hiyoshi 3-1-14, Kohoku, Yokohama, Japan
}


\begin{abstract}

Recently the complete characterization of a general Gaussian 
dissipative system having a unique pure steady state was obtained 
in [Koga $\&$ Yamamoto 2012, Phys. Rev. A 85, 022103]. 
This result provides a clear guideline for engineering an 
environment such that the dissipative system has a desired 
pure steady state such as a cluster state. 
In this paper, we describe the system in terms of a quantum 
stochastic differential equation (QSDE) so that the environment 
channels can be explicitly dealt with. 
Then a physical meaning of that characterization, which 
cannot be seen without the QSDE representation, is clarified; 
more specifically, the nullifier dynamics of any Gaussian system 
generating a unique pure steady state is passive. 
In addition, again based on the QSDE framework, we provide a 
general and practical method to implement a desired dissipative 
Gaussian system, which has a structure of quantum state transfer. 

\end{abstract}

\maketitle


\section{Introduction}

Towards quantum state preparation, which clearly plays a key 
part in quantum information processing, recently several 
dissipation-based approaches have been proposed. 
The basic idea of those approaches originates from the trivial 
fact that a thermal environment drives any state to the stable 
ground state. 
However, it has been shown that we can sometimes {\it engineer} 
a desired dissipative environment such that the corresponding 
stable state is a nontrivial and useful one, e.g., a highly 
entangled pure state 
\cite{Poyatos,Yamamoto,Cirac1,Kraus,TicozziViola2008,Kraus_nature,
VerstraeteNaturePhys,Vollbrecht,Ficek2009,Schirmer,Cirac2,Polzik,
KogaYamamoto,TicozziViola2011}. 
More specifically, under some conditions we are allowed to 
synthesize an open quantum system described by the Markovian 
master equation 
\begin{equation}
\label{me}
   \frac{d\hat \rho_t}{dt} 
    = - i[\hat H , \hat\rho_t] 
      + \sum_{k=1}^m \Big( 
          \hat L_k \hat\rho_t \hat L_k^\dagger 
           -\frac{1}{2} \hat L_k^\dagger \hat L_k \hat\rho_t 
            -\frac{1}{2} \hat\rho_t \hat L_k^\dagger \hat L_k \Big),
\end{equation}
such that $\hat\rho_t$ must converge into a given desired pure 
state $\hat \rho_\infty$; 
that is, the Hamiltonian $\hat H$ and the dissipative channel 
$\hat L_k~(k=1,\ldots,m)$ are appropriately synthesized to 
achieve this goal. 
One of the main advantages of this approach is that the target 
state $\hat\rho_\infty$ is clearly robust against any perturbation 
to the state $\hat\rho_t$ during the dynamical process. 
In particular, it is independent on the initial state preparation.

In the finite dimensional case a necessary and sufficient 
condition for Eq. \eqref{me} to have a pure steady state was 
obtained in \cite{Yamamoto,Kraus}, and especially in \cite{Kraus} 
the authors provided a sufficient condition for $\hat\rho_\infty$ 
to be unique. 
The uniqueness characterization is of particular importance, 
because without such condition the desired convergence into the 
target state cannot be guaranteed. 
For infinite-dimensional systems, on the other hand, in 
\cite{KogaYamamoto} the author particularly focused on a general 
Gaussian dissipative system and provided a complete 
parameterization of the system having a unique pure steady state. 
The merit of focusing on the class of Gaussian systems lies not 
only in its importance in quantum information technologies 
\cite{Braunstein2005,Furusawa2011} but also in the fact that 
the parameterization is obtained in an easily-tractable manner 
in the phase space; 
actually the uniqueness of $\hat\rho_\infty$ can be readily 
checked by simply calculating the rank of a specific matrix, 
while in the finite-dimensional case \cite{Kraus} we are 
required to verify that there is no specific subspace in the 
Hilbert space.

In this paper, we study a Gaussian system having a unique 
pure steady state in terms of a {\it quantum stochastic 
differential equation} (QSDE) \cite{Hudson,Belavkin,Gardiner,Wiseman}. 
The use of a QSDE allows us to describe the dynamics of an open 
system in a form where the stochastic environment channels appear 
explicitly. 
The master equation \eqref{me} is obtained as a result of averaging 
out all such stochastic effects brought from the environment.

The contribution of this paper is twofold. 
The first one is that we clarify the physical meaning of the 
conditions for the Gaussian system to have a unique pure steady 
state, which cannot be clearly seen when dealing with only the 
master equation. 
This result is obtained through investigating the QSDE of the 
corresponding (complex) {\it nullifier}. 
In general, it is known that any pure Gaussian state can be 
characterized as the common zero eigenstate of the corresponding 
nullifier operators \cite{Menicucci2011}; 
this is the reason why investigating the full behavior of the 
nullifier provides new information about the dynamic process 
towards the target pure state. 
Actually, we show that the nullifier dynamics of any Gaussian 
system generating a unique pure steady state is {\it passive}. 
As a byproduct, the result is used to show a certain trade-off 
between the closeness of the steady state to a target 
{\it Gaussian cluster state} 
\cite{Zhang2006,Menicucci2006,vanLoock2007,Menicucci2007} and 
the convergence time into that steady state.

In the previous result \cite{KogaYamamoto}, although the 
mathematical characterization of the desired dissipative 
channel $\hat L_k$ was obtained, its actual implementation 
was not discussed. 
Actually, the resulting desired dissipative channels usually 
have to non-locally act on the system, and no general method 
to effectively implement such dissipative channels is known. 
The second contribution of this paper is to give a partial 
answer to the question of how to practically construct a 
desired dissipative system. 
The proposed scheme has a structure of {\it quantum state 
transfer} from light to a matter \cite{Parkins1999,Zhang2003,
Julsgaard2004}; 
more specifically, a desired state of light is first generated 
and then that light field interacts with the oscillator system 
(memory), which as a result acquires the desired state by 
dissipation. 
This scheme is indeed practical, because, as shown in 
\cite{Menicucci2007}, any pure Gaussian cluster state of light 
can be effectively generated using some beam splitters and OPOs. 
Note that the QSDE approach actually has to be taken in order to 
explicitly describe the input light field.

We use the following notations: 
for a matrix $A=(a_{ij})$, the symbols $A^\dagger$, $A^\top$, 
and $A^\sharp$ represent its Hermitian conjugate, transpose, and 
elementwise complex conjugate of $A$, i.e., 
$A^\dagger=(a_{ji}^*)$, $A^\top=(a_{ji})$, and 
$A^\sharp=(a_{ij}^*)=(A\dgg)^\top$, respectively. 
For a matrix of operators, $\hat A=(\hat a_{ij})$, we use the 
same notation, in which case $\hat a_{ij}^*$ denotes the adjoint 
to $\hat a_{ij}$. 
$I_n$ denotes the $n\times n$ identity matrix. 
$\Re$ and $\Im$ denote the real and imaginary parts, respectively.


\section{Preliminaries}

In this section, a brief introduction to a Gaussian system 
and its QSDE representation is given. 
Then we review the result of \cite{KogaYamamoto}.


\subsection{Gaussian dissipative systems}

A general $n$-mode bosonic system consists of $n$ subsystems 
with canonical conjugate pairs $(\hat q_i,~\hat p_i)$. 
Denote the vector of total system variables by 
$\hat x :=(\hat q_1, \ldots, \hat q_n, \hat p_1, \ldots, \hat p_n)^\top$. 
The canonical commutation relation $[\hat q_i,~\hat p_j]=i \delta _{ij}$ 
then leads to 
\begin{eqnarray*}
   \hat x \hat x ^\top -(\hat x \hat x^\top )^\top 
      = i \Sigma,~~
   \Sigma = \left(\begin{array}{cc}
               0 & I_n \\
               -I_n & 0
            \end{array}\right). 
\end{eqnarray*}
Now let $\hat \rho$ be the density operator of this system and 
write the mean vector by $\mean{\hat x}$ and the covariance matrix 
by $V=\mean{\Delta \hat x \Delta \hat x^\top 
 + (\Delta \hat x \Delta \hat x^\top)^\top}/2,~
\Delta \hat x=\hat x-\mean{\hat x}$, where the mean 
$\mean{\hat X}=\Tr(\hat X\hat \rho)$ is taken elementwise. 
Note that the uncertainty relation $V+i\Sigma/2 \geq 0$ holds. 
A Gaussian state can be characterized by only the mean vector 
and the covariance matrix. 
A particularly important fact is that the covariance matrix 
$V$ corresponding to a pure Gaussian state always has the 
following general representation \cite{Menicucci2011,Simon}: 
\begin{equation}
\label{pure CM general}
    V=\frac{1}{2}SS^\top,~~~
    S = \left( \begin{array}{cc} 
            Y^{-1/2}  & 0 \\
            XY^{-1/2} & Y^{1/2} \\
        \end{array}\right), 
\end{equation}
where $X$ and $Y$ are $n\times n$ real symmetric and real 
positive definite matrices (i.e., $Y=Y^\top>0$), respectively. 
In other words, a pure Gaussian state is completely parameterized 
by $X$ and $Y$. 
An important merit of this representation is that the complex 
{\it graph matrix} $Z:=X+iY$ can be used for a graphical calculus 
for several Gaussian pure states \cite{Menicucci2011}. 
In particular, a pure Gaussian state $\ket{\psi_Z}$ having 
the covariance matrix \eqref{pure CM general} always satisfies 
\begin{equation}
\label{nullifier}
    \hat r \ket{\psi_Z}=0,~~~
    \hat r:=(-Z,~I_n)\hat x
           = \left( \begin{array}{c} 
               \hat p_1 \\
               \vdots   \\
               \hat p_n \\
             \end{array}\right)
           -Z\left( \begin{array}{c} 
               \hat q_1 \\
               \vdots   \\
               \hat q_n \\
              \end{array}\right), 
\end{equation}
where the equation means that each entry of $\hat r$ acts 
on $\ket{\psi_Z}$. 
Conversely, if a pure Gaussian state $\ket{\psi}$ satisfies 
$(-Z,~I_n)\hat x\ket{\psi}=0$, then we have $\ket{\psi}=\ket{\psi_Z}$. 
The vector of operators $\hat r$ is called the {\it nullifier} 
for the pure Gaussian state $\ket{\psi_Z}$.

A linear system is such that the Hamiltonian $\hat H$ and $k$-th 
dissipative channel $\hat L_k$ in Eq. \eqref{me} are respectively 
characterized by 
\begin{equation}
\label{linear H and L}
     \hat H = \frac{1}{2} \hat x^\top G \hat x,~~
     \hat L_k = c_k^\top \hat x, 
\end{equation}
where $G=G^\top \in \textbf{R}^{2n \times 2n}$ and 
$c_k \in \textbf{C}^{2n}$. 
For this system, the time-evolution of $\mean{\hat x_t}$ and $V_t$ 
with the state $\hat\rho_t$ obeying Eq. \eqref{me} are given by 
$d\mean{\hat x_t}/dt = A \mean{\hat x_t}$ and 
$dV_t/dt = AV_t+V_tA^\top + D$, respectively. 
Here, $A=\Sigma[G+\Im(C^\dagger C)]$ and 
$D=\Sigma\Re(C^\dagger C)\Sigma^\top$ with 
$C=(c_1,\ldots,c_m)^\top\in \textbf{C}^{m\times 2n}$ 
(see \cite{Wiseman} for more detailed discussion). 
In this paper, we assume that the initial state of the dynamics is 
Gaussian; 
then, at any given time $t$ the state is also Gaussian with mean 
$\mean{\hat x_t}$ and covariance $V_t$, hence let us call such a 
linear system the Gaussian system. 
A steady state of the Gaussian system exists only when $A$ is a 
{\it Hurwitz} matrix, i.e., all the eigenvalues of $A$ have 
negative real parts. 
If it exists, the mean vector is $\mean{\hat x_\infty}=0$ and 
the covariance matrix $V_\infty$ is given by the unique 
solution to the following matrix equation: 
\begin{eqnarray}
\label{ale}
   AV_\infty + V_\infty A^\top + D = 0. 
\end{eqnarray} 
%


\subsection{The QSDE framework}

The situation we have in mind is that the system interacts 
with countable set of environment channels. 
The time-evolution of an observable of this open system is 
described in terms of a QSDE. 
A most simple form of this equation is obtained when the 
environment channels are all independent vacuum fields with 
ideal Markovian approximation taken. 
Let $\hat a_i(t)$ be the annihilation operator of the $i$-th 
vacuum field; 
then the Markovian approximation means that $\hat a_i(t)$ 
instantaneously interacts with the system and satisfies 
the CCR $[\hat a_i(s), \hat a_j^*(t)]=\delta_{ij}\delta(t-s)$. 
Define the field annihilation process operator by 
$\hat A_i(t)=\int_0^t \hat a_i(s)ds$, then this CCR leads to 
the following {\it quantum Ito rule}: 
\begin{equation}
\label{Ito rule}
    d\hat A_i d\hat A_j^*=\delta_{ij}dt,~~
    d\hat A_i d\hat A_j=d\hat A_i^* d\hat A_j^*
                       =d\hat A_i d\hat A_j^*=0.
\end{equation}
The system-field coupling in the time interval $[t, t+dt)$ is 
described by the unitary operation 
$\hat U(t+dt, t)
={\rm exp}[\sum_i(\hat L_i d\hat A_i^* - \hat L_i^* d\hat A_i)]$, 
where $\hat L_i$ is the system operator representing the coupling 
with the $i$-th vacuum field. 
Then the system observable at time $t$, 
$j_t(\hat X)=\hat U_t^*\hat X \hat U_t$, obeys the Ito-type QSDE
\begin{eqnarray}
& & \hspace*{-1em}
\label{QSDE}
    dj_t(\hat X)=j_t\Big( i[\hat H, \hat X] 
       + \sum_{i=1}^m \big( 
         \hat L_i^* \hat X \hat L_i
          -\frac{1}{2} \hat L_i^* \hat L_i \hat X
           -\frac{1}{2} \hat X \hat L_i^* \hat L_i \big) \Big)dt
\nonumber \\ & & \hspace*{5em}
    \mbox{}
      + \sum_{i=1}^m \Big( 
        j_t([\hat X, \hat L_i])d\hat A_i^*
          - j_t([\hat X, \hat L_i^*])d\hat A_i \Big), 
\end{eqnarray}
where an additional system Hamiltonian $\hat H$ has been 
added. 
Note that $\hat U_{t+dt}=\hat U(t,t+dt)\hat U_t$. 
The mean value $\mean{j_t(\hat X)}$ is represented using the 
(unconditional) density operator $\hat\rho_t$ by 
$\mean{j_t(\hat X)}=\Tr(\hat X\hat \rho_t)$, which leads to the 
master equation \eqref{me}. 
The change of the field operator can also be dealt with explicitly; 
the output field $\hat A'_i:=j_t(\hat A_i)$ after the interaction 
satisfies 
\begin{equation}
\label{general output}
    d\hat A'_i = j_t(\hat L_i)dt + d\hat A_i.
\end{equation}

We are interested in the QSDE whose system Hamiltonian and 
dissipative channels are given by Eq. \eqref{linear H and L}. 
Let us define 
$\hat {\cal A}_t = (\hat A_1, \ldots, \hat A_m)^\top$, then the 
vector of system quadratures 
$\hat x_t=(j_t(\hat q_1), \ldots, j_t(\hat q_n), 
           j_t(\hat p_1), \ldots, j_t(\hat p_n) )^\top$ 
satisfies the linear QSDE \cite{Wiseman,GoughPRA2008,
GoughPRA2009,GoughPRA2010,Nurdin2010,Petersen2011,Petersen2012}:
\begin{equation}
\label{linear QSDE}
   d\hat x_t= A\hat x_t dt - i\Sigma C^\dagger d\hat{\cal A}_t
                           + i\Sigma C^\top d\hat{\cal A}_t^\sharp, 
\end{equation}
where the system matrices $A$ and $C$ were defined in 
Section 2 (a). 
It is easy to see that the linear QSDE \eqref{linear QSDE} 
actually leads to the time-evolutions of the mean and the 
covariance matrix: 
$d\mean{\hat x_t}/dt = A \mean{\hat x_t}$ and 
$dV_t/dt = AV_t+V_tA^\top + D$. 
Also the output field equation \eqref{general output} of 
$\hat{\cal A}'_t = (\hat A'_1, \ldots, \hat A'_m)^\top$ then 
becomes
\begin{equation}
\label{linear output}
   d\hat{\cal A}'_t= C\hat x_t dt + d\hat{\cal A}_t. 
\end{equation}
%


\subsection{The dissipative Gaussian system generating a 
pure steady state}

In \cite{KogaYamamoto}, some conditions for a dissipative 
Gaussian system to have a unique pure steady state were 
obtained. 
A particularly useful result from the environment engineering 
viewpoint is the following (recall $Z:=X+iY$):

{\it Theorem 1} \cite{KogaYamamoto}: 
Let $V$ be a given covariance matrix of the form 
\eqref{pure CM general}. 
Then, this is the unique solution of Eq. \eqref{ale} 
if and only if the system matrices are represented by
\begin{eqnarray}
& & \hspace*{-1em}
\label{C representation}
    C = P^\top (-Z, I_n), 
\\ & & \hspace*{-1em}
\label{G representation}
    G=\left( \begin{array}{cc} 
          XRX + YRY - \Gamma Y^{-1}X - XY^{-1}\Gamma^\top~~ 
                          & -XR + \Gamma Y^{-1} \\
          -RX + Y^{-1}\Gamma^\top & R        \\
      \end{array}\right), 
\end{eqnarray}
where $P$ is a complex $n\times m$ matrix, $R$ is a real 
$n\times n$ symmetric matrix, and $\Gamma$ is a real $n\times n$ 
skew symmetric matrix (i.e., $\Gamma+\Gamma^\top=0$), and moreover, 
$P$ and $Q:=-iRY-Y^{-1}\Gamma^\top$ satisfy the following rank 
condition: 
\begin{equation}
\label{rank condition}
   {\rm rank}(P,~QP, \ldots,~Q^{n-1}P)=n. 
\end{equation}

This theorem states that any dissipative linear system having 
a unique pure Gaussian steady state is completely parameterized 
by the three matrices $P, R$, and $\Gamma$, which further have to 
satisfy the rank condition \eqref{rank condition}. 
In \cite{KogaYamamoto}, this result was obtained through a 
fully algebraic treatment of Eq. \eqref{ale}, and the physical 
meanings of the conditions were not discussed. 
As mentioned in Section 1, nevertheless, they will be clarified 
within the QSDE framework; 
for convenience of the later discussion, we note that $G$ satisfies 
\begin{equation}
\label{G condition}
    G\Sigma^\top
           \left( \begin{array}{c} 
             -Z  \\
             I_n \\
           \end{array}\right)
          =\left( \begin{array}{c} 
             -Z  \\
             I_n \\
           \end{array}\right)Q. 
\end{equation}
%


\section{Dynamics of the nullifier}

As seen in Eq. \eqref{nullifier}, the pure state 
$\ket{\psi_Z}$ is the common zero-eigenstate of the nullifier 
vector $\hat r=(-Z, I_n)\hat x$. 
Hence it is worth to see the time-evolution of $\hat r_t$, 
when the conditions shown in Theorem 1 are satisfied. 
Noting Eqs. \eqref{C representation} and \eqref{G condition}, 
we have 
\begin{eqnarray}
& & \hspace*{-1.5em}
   (-Z, I)A=(-Z, I)\Sigma G 
             + \frac{1}{2i} (-Z, I)\Sigma 
                  (C^\dagger C - C^\top C^\sharp)
\nonumber \\ & & \hspace*{-1em}
   =Q^\top (-Z, I) + \frac{1}{2i} (-I, -Z) 
      \Big\{ \left( \begin{array}{c} 
              -Z^\sharp \\
              I  \\
             \end{array}\right)P^\sharp P^\top (-Z, I)
           - \left( \begin{array}{c} 
              -Z \\
              I  \\
             \end{array}\right)PP^\dagger (-Z^\sharp, I)  \Big\}
\nonumber \\ & & \hspace*{-1em}
   =Q^\top (-Z, I) + \frac{Z^\sharp -Z}{2i}P^\sharp P^\top (-Z, I)
   =(Q^\top - YP^\sharp P^\top) (-Z, I),
\nonumber \\ & & \hspace*{-1.5em}
   (-Z, I)(-i\Sigma C^\dagger)
     = -i(-Z, I)\Sigma 
           \left( \begin{array}{c} 
              -Z^\sharp \\
              I  \\
           \end{array}\right)P^\sharp
     =i(Z-Z^\sharp)P^\sharp = -2YP^\sharp, 
\nonumber \\ & & \hspace*{-1.5em}
   (-Z, I)(i\Sigma C^\top)
     = i(-Z, I)\Sigma 
           \left( \begin{array}{c} 
              -Z \\
              I  \\
           \end{array}\right)P
     =0. 
\nonumber
\end{eqnarray}
Therefore, multiplying both sides of Eq. \eqref{linear QSDE} by 
$(-Z, I_n)$ from the left, we have 
\begin{equation}
\label{nullifier dynamics}
    d\hat r_t=(Q^\top - YP^\sharp P^\top)\hat r_t dt 
                 -2YP^\sharp d\hat {\cal A}_t. 
\end{equation}
Regarding the output process \eqref{linear output}, as 
$C\hat x_t=P^\top(-Z, I_n)\hat x_t=P^\top\hat r_t$, it is written 
by 
\begin{equation}
\label{nullifier output}
    d\hat{\cal A}'_t=P^\top\hat r_t dt + d\hat {\cal A}_t. 
\end{equation}
The coefficient matrix of the dynamics of $\hat r$ has the 
following property.

{\it Proposition 2}: 
The matrix $Q^\top - YP^\sharp P^\top$ is Hurwitz if and only if 
the rank condition \eqref{rank condition} is satisfied.

{\it Proof}: 
Let $b\in{\mathbb C}^n$ and $\lambda\in{\mathbb C}$ be the 
eigenvector and the eigenvalue of $Q^\top - YP^\sharp P^\top$, 
respectively; 
i.e., $(Q^\top - YP^\sharp P^\top)b=\lambda b$. 
Then, multiplying this equation by $b^\dagger Y^{-1}$ from the left, 
we have 
\[
    b^\dagger(Y^{-1}Q^\top - P^\sharp P^\top)b
      =\lambda \|Y^{-1/2}b\|^2, 
\]
where $\|\bullet\|$ denotes the standard Euclidean norm. 
This immediately yields 
$b^\dagger(Q^\sharp Y^{-1} - P^\sharp P^\top)b
=\lambda^* \|Y^{-1/2}b\|^2$. 
Recall now that $Q=-iRY-Y^{-1}\Gamma^\top$, hence 
$QY^{-1}=-iR-Y^{-1}\Gamma^\top Y^{-1}$ is skew Hermitian. 
Therefore, adding the above two equations yields 
$-2\|P^\top b\|^2=(\lambda+\lambda^*)\|Y^{-1/2}b\|^2$, and 
we have $\Re(\lambda)=-\|P^\top b\|^2/\|Y^{-1/2}b\|^2$. 
Let us here assume that $P^\top b=0$. 
Then, we have $Q^\top b=\lambda b$, and 
the matrix ${\cal C}:=(P, QP, \ldots, Q^{n-1}P)$ satisfies 
\[
    {\cal C}^\top b
     =\left( \begin{array}{c} 
               P^\top b \\
               P^\top Q^\top b  \\
               \vdots \\
               P^\top (Q^\top)^{n-1} b  \\
             \end{array}\right)
     =\left( \begin{array}{c} 
               P^\top b \\
               \lambda P^\top b  \\
               \vdots \\
               \lambda^{n-1} P^\top b  \\
             \end{array}\right)
     =0. 
\]
But this is contradiction to the assumption \eqref{rank condition}, 
thus $P^\top b\neq 0$. 
As a result, $\Re(\lambda)$ is strictly negative, implying that 
the matrix $Q^\top - YP^\sharp P^\top$ is Hurwitz.

On the other hand, if ${\cal C}$ is not of rank $n$, there 
exists an eigenvector of $Q^\top - YP^\sharp P^\top$, say $b_0$, 
that satisfies $P^\top b_0=0$. 
Then, from the above discussion, the corresponding eigenvalue 
$\lambda$ satisfies $\Re(\lambda)=0$, hence $Q^\top-YP^\sharp P^\top$ 
is not Hurwitz. 
\hfill $\blacksquare$

Based on the above result, we can verify that the target 
pure Gaussian state is certainly generated. 
To see this, let us multiply all the entries of the nullifier 
dynamics \eqref{nullifier dynamics} by the system-field composite 
state vector $\ket{\Psi}=\ket{\psi}\otimes\ket{0}$ from the right 
($\ket{0}$ is the vacuum state). 
Then, due to the relation $d\hat{\cal A}_t\ket{0}=0$, we have 
\[
    \frac{d}{dt}\hat r_t\ket{\Psi} 
      = (Q^\top - YP^\sharp P^\top)\hat r_t\ket{\Psi}. 
\]
The Hurwitz property of the matrix $Q^\top - YP^\sharp P^\top$ 
is equivalent to asymptotic stability of the dynamics, thus the 
nullifier vector 
$\hat r_t\ket{\Psi}=\hat U_t^* \hat r \hat U_t\ket{\Psi}$ 
converges to zero. 
Therefore, in the Schr\"odinger picture, 
$\hat U_\infty\ket{\Psi}$ is the common zero-eigenstate of 
$\hat r$, meaning that the system state becomes the target pure 
Gaussian state $\ket{\psi_Z}$ as $t\rightarrow\infty$.

The Hurwitz property obtained above allows us to obtain a 
specific input-output relation from the incoming field 
$\hat{\cal A}_t$ to the outgoing field $\hat{\cal A}'_t$, 
through Eqs. \eqref{nullifier dynamics} and \eqref{nullifier output}. 
For this purpose, it is convenient to move into the frequency 
domain where both of these field operators as well as the internal 
system variable $\hat x_t$ are all Fourier transformed. 
The Fourier transformation of Eqs. \eqref{nullifier dynamics} 
and \eqref{nullifier output} are given by 
\begin{eqnarray}
& & \hspace*{0em}
    i\omega \tilde{r}(\omega)=
      (Q^\top - YP^\sharp P^\top)\tilde{r}(\omega)
        - 2YP^\sharp \tilde{\cal A}(\omega), 
\nonumber \\ & & \hspace*{0em}
    \tilde{\cal A}'(\omega)=P^\top \tilde{r}(\omega) 
        + \tilde{\cal A}(\omega), 
\nonumber
\end{eqnarray}
where the tilde notation denotes the Fourier transformed operator. 
Note that for instance $\tilde{A}_i(\omega)$ is not the Fourier 
transformation of $\hat A_i(t)$ but $\hat a_i(t)$. 
Also, more precisely, we should take Laplace transformation 
$\hat a_i(t) \rightarrow \tilde{A}_i(s)$ and set $s=+0+i\omega$ 
to obtain the Fourier transformation; 
for the rigorous treatment, see 
\cite{GoughPRA2008,GoughPRA2009,GoughPRA2010}. 
As a result, we have the input-output map from 
$\tilde{\cal A}(\omega)$ to $\tilde{\cal A}'(\omega)$: 
\begin{equation}
\label{input output map}
   \tilde{\cal A}'(\omega)=
      F(\omega)\tilde{\cal A}(\omega), ~~~
   F(\omega):=I_m
     -2P^\top (i\omega - Q^\top + YP^\sharp P^\top)^{-1}YP^\sharp. 
\end{equation}
Here we have used the Hurwitz property to justify that the 
initial contribution of the system was ignored in 
Eq. \eqref{input output map}. 
The $m\times m$ matrix $F(\omega)$, called the {\it transfer 
function matrix}, has a striking property as shown below.

{\it Proposition 3}: 
The transfer function matrix $F(\omega)$ is unitary for all $\omega$.

{\it Proof}: 
The proof directly follows from Lemma 2 of \cite{GoughPRA2008}, 
but here it is given for convenience of readers. 
First, to simplify the calculation, let us define 
$\bar{P}=Y^{1/2}P$ and $\bar{Q}=Y^{1/2}QY^{-1/2}$. 
As $Q=-iRY-Y^{-1}\Gamma^\top$, $\bar{Q}$ is skew Hermitian; 
$\bar{Q}+\bar{Q}^\dagger=0$. 
With this notation, the transfer function matrix is represented 
by 
$F(\omega)
:=I_m-2\bar{P}^\top
(i\omega - \bar{Q}^\top + \bar{P}^\sharp \bar{P}^\top)^{-1}
\bar{P}^\sharp$. 
Therefore, we have
\begin{eqnarray}
& & \hspace*{-1em}
\label{unitarity proof}
   F(\omega)^\dagger F(\omega)
    = I_m - 2\bar{P}^\top
          (-i\omega - \bar{Q}^\sharp 
               + \bar{P}^\sharp\bar{P}^\top)^{-1}\bar{P}^\sharp
        - 2\bar{P}^\top
          (i\omega - \bar{Q}^\top 
               + \bar{P}^\sharp\bar{P}^\top)^{-1}\bar{P}^\sharp
\nonumber \\ & & \hspace*{1em}
   \mbox{}
   + 2\bar{P}^\top
          (-i\omega - \bar{Q}^\sharp 
              + \bar{P}^\sharp \bar{P}^\top)^{-1}
                  (2\bar{P}^\sharp \bar{P}^\top)
          (i\omega - \bar{Q}^\top 
              + \bar{P}^\sharp\bar{P}^\top)^{-1}\bar{P}^\sharp. 
\end{eqnarray}
Here it follows from $\bar{Q}+\bar{Q}^\dagger=0$ that
\[
    2\bar{P}^\sharp \bar{P}^\top
     = (-i\omega - \bar{Q}^\sharp + \bar{P}^\sharp\bar{P}^\top)
       + (i\omega - \bar{Q}^\top + \bar{P}^\sharp\bar{P}^\top). 
\]
Substituting this expression for the last term of 
Eq. \eqref{unitarity proof}, we end up with the relation 
$F(\omega)^\dagger F(\omega)=I_m$, hence $F(\omega)$ is unitary 
for all $\omega$. 
\hfill $\blacksquare$

This result states that the output power spectrum is flat in 
all the frequency domain, i.e., 
$\mean{\tilde{\cal A}'(\omega)\tilde{\cal A}'(\omega)^\dagger}
=F(\omega)\mean{\tilde{\cal A}(\omega)
\tilde{\cal A}(\omega)^\dagger}F(\omega)^\dagger=I_m$. 
This means that, at steady state, it is impossible to extract 
any information about the internal system, as long as the matrix 
$Q^\top - YP^\sharp P^\top$ is Hurwitz and the input fields are 
in vacuum or coherent states.

Now we arrive at the stage where the physical meanings of the 
conditions given in Theorem~1 can be clarified. 
First of all, the structure of Eqs. \eqref{nullifier dynamics} 
and \eqref{nullifier output} as well as the above two propositions 
remind us that the dynamics of the nullifier is a generalization 
of that for the simple single-mode optical damped cavity whose 
QSDE is described by 
\begin{equation}
\label{damped cavity}
     d\hat a_t=\big(i\Delta-\frac{\kappa}{2}\big)\hat a_t dt 
                -\sqrt{\kappa}d\hat A_t,~~~
     d\hat A'_t=\sqrt{\kappa}\hat a_t dt + d\hat A_t, 
\end{equation}
where $\hat a_t$ and $\hat A_t$ denote the intra-cavity mode 
and the incoming vacuum field mode, respectively. 
$\Delta$ and $\kappa$ denote the detuning and the damping rate, 
respectively. 
Clearly, the state evolves into the vacuum, and also we have 
$\mean{\tilde{A}'(\omega)^*\tilde{A}'(\omega)}
=\mean{\tilde{A}(\omega)^*\tilde{A}(\omega)}=1$ 
for all $\omega$, as in the nullifier case. 
These properties arise due to 
(i) that energy is not supplied through the Hamiltonian, 
(ii) that the field does not supply energy but simply brings about 
the damping of the system, 
and (iii) that the system is asymptotically stable. 
Mathematically, the first two statements mean that the dynamics 
does not contain the creation operators $\hat a^*_t$ and 
$\hat A^*_t$. 
Actually, regarding the first one, if the cavity contains 
a degenerate parametric amplifier, which is described by the 
Hamiltonian $\hat H_{DPA}=i(\hat a^*\mbox{}^2-\hat a^2)$, 
then the QSDE needs to be described in terms of both $\hat a$ 
and $\hat a^*$. 
The last condition (iii) guarantees that the state uniquely 
converges into the vacuum as well as that the output field does 
not contain any information about the system at steady state. 
Systems having the properties (i)-(iii) are in general called 
{\it passive systems} 
\cite{GoughPRA2008,Nurdin2010,Petersen2011,Petersen2012}.

The above discussion implies that the nullifier dynamics is 
passive; 
more precisely, we obtain the physical meanings of the 
conditions \eqref{C representation}, \eqref{G representation}, and 
\eqref{rank condition} as follows. 
\begin{itemize}

\item
The matrix $C$ has the form given in Eq. \eqref{C representation} 
so that the creation process $\hat{\cal A}^\sharp_t$ does not appear 
in the QSDE of $\hat r_t$; 
as mentioned above, this is equivalent to that there is no energy 
supply from the environment to the nullifier. 

\item
The matrix $G$ has the form given in Eq. \eqref{G representation} 
so that the corresponding Hamiltonian $\hat H=\hat x^\top G \hat x/2$ 
does not supply energy for the nullifier. 

\item
The rank condition \eqref{rank condition} implies that the 
coefficient matrix $Q^\top - YP^\sharp P^\top$ is Hurwitz, or 
equivalently the asymptotic stability of the dynamics of 
the nullifier. 
This guarantees that the output power spectrum is flat in all 
frequencies, meaning that the output field does not contain any 
information about the system at steady state. 

\end{itemize}

The last statement can be understood by studying the {\it filtering 
equation} \cite{Belavkin,Bouten}, which enables us to update the 
conditional state based on the measurement result of the output 
field. 
In general, when the state of the master equation reaches the 
steady state and it is pure, then the corresponding filtering 
equation is identical to the master equation, meaning that we do 
not obtain any new information through measuring the output field 
for updating our knowledge. 
Note that this does not mean that the system is not controllable.


\section{Quantum state transfer for the dissipative system 
engineering}

We have seen in Theorem 1 how the dissipative channel 
$\hat L_k =c_k^\top \hat x$ with $C=(c_1^\top, \ldots, c_m^\top)$ 
should be chosen to engineer a desired Gaussian dissipative system. 
When we aim to generate a certain (useful) Gaussian state, however, 
it often turns out that the resulting $\hat L_k$ has to possess a 
specific structure which is hard to actually implement. 
For instance, a dissipative channel interacting with all the nodes, 
i.e., 
$\hat L=\ell_1\hat q_1 + \ell_2\hat q_2 + \ldots + \ell_n\hat q_n$, 
will be hard to implement. 
In this section, for the specific case where the system is 
subjected to $n$ independent input optical fields (i.e., $m=n$), 
we provide a practical procedure for implementing desired dissipative 
channels.

Let us first introduce the field quadratures ($k=1, \ldots, n$)
\begin{equation}
    \hat Q_k=(\hat A_k + \hat A_k^*)/\sqrt{2},~~~
    \hat P_k=(\hat A_k - \hat A_k^*)/\sqrt{2}i,~~~
\end{equation}
which satisfy the canonical commutation relation 
$[d\hat Q_i, d\hat P_j]=\delta_{ij}dt$. 
Then, defining $\hat {\cal Q}=(\hat Q_1, \ldots, \hat Q_n)^\top$ 
and $\hat {\cal P}=(\hat P_1, \ldots, \hat P_n)^\top$, we find 
that the QSDE \eqref{linear QSDE} is rewritten by 
\begin{equation}
\label{linear QSDE quadrature}
   d\hat x_t= A\hat x_t dt 
         + \sqrt{2}\Sigma (C_r^\top, C_i^\top)d\hat{\cal W}_t,~~~
   \hat{\cal W}_t
         =\left( \begin{array}{c} 
                   \hat {\cal Q}_t  \\
                   \hat {\cal P}_t  \\
                 \end{array}\right), 
\end{equation}
where $C_r=\Re(C)$ and $C_i=\Im(C)$.

\begin{figure}[t]
\centering
\begin{picture}(200,90)
\put(0,0)
{\includegraphics[width=2.5in]{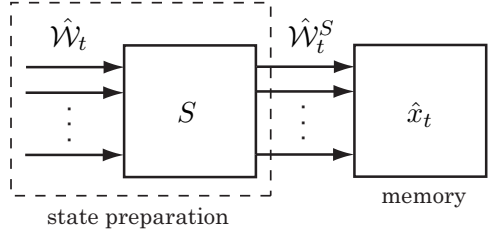}}

\put(15,70){$\hat{\cal W}_t$}
\put(105,70){$\hat{\cal W}_t^S$}
\put(63,41){$S$}
\put(150,41){$\hat{x}_t$}

\end{picture}
\caption{
Quantum state transfer from the input mode $\hat{\cal W}_t^S$ 
to the memory mode $\hat x_t$. 
}
\end{figure}

The situation we have in mind is that a desired pure Gaussian 
state of light is first generated, and then, that state is 
transferred to the system through the system-field coupling; 
see Figure 1. 
This is the framework of the quantum state transfer 
\cite{Parkins1999,Zhang2003,Julsgaard2004}; 
in this case the system is called the {\it memory} and it should 
be independent on the input state we will transfer. 
More specifically, the target mode $\hat{\cal W}^S_t$ is obtained 
from the vacuum mode $\hat{\cal W}_t$ through the transformation 
$\hat{\cal W}^S_t=S\hat{\cal W}_t$ where the symplectic matrix 
$S$ is given in Eq. \eqref{pure CM general}; 
then the quantum Ito rule \eqref{Ito rule} gives 
\[
    d\hat{\cal W}^S_t (d\hat{\cal W}^S_t)^\top
     = S d\hat{\cal W}_t d\hat{\cal W}_t^\top S^\top
     = S (I_n + i\Sigma) S^\top dt/2
     = (SS^\top/2 + i\Sigma/2)dt, 
\]
implying that the covariance matrix of the input field 
$\hat{\cal W}^S_t$ is certainly $V=SS^\top/2$ (in the rigorous 
sense this statement should be given in terms of the power 
spectrum density; see \cite{GoughPRA2009}). 
Note that $S$ can contain a squeezing process, which was not 
included in the original QSDE framework of Hudson and Parthasarathy 
\cite{Hudson}; see \cite{GoughPRA2010} for a detailed discussion. 
Now the system \eqref{linear QSDE quadrature} is written as 
\begin{equation}
\label{linear QSDE transfer}
   d\hat x_t= A\hat x_t dt + B d\hat{\cal W}^S_t,~~~
   B:=\sqrt{2}\Sigma (C_r^\top, C_i^\top)S^{-1},
\end{equation}
where, as shown above, the new input field $\hat{\cal W}^S_t$ 
carries information of the target Gaussian state. 
The system, which serves as a memory, should satisfy the following 
two requirements:
\begin{itemize}

\item[(R1)]
The memory system \eqref{linear QSDE transfer} should not 
possess any information about the input state; 
that is, the system's coefficient matrices $A$ and $B$ 
should be independent on $Z=X+iY$. 

\item[(R2)]
The state of the memory system \eqref{linear QSDE transfer} 
should converge to the target Gaussian state with covariance 
matrix $V_\infty=SS^\top/2$. 
That is, $C$ and $G$ should be of the form \eqref{C representation} 
and \eqref{G representation} with $P$ and $Q$ satisfying the rank 
condition \eqref{rank condition}. 

\end{itemize}
Below we give a characterization of the desired memory system:

{\it Proposition 4}: 
Assume that the system satisfies the requirements (R1) and (R2). 
Then, the system has to be of the form 
\begin{equation}
   d\hat x_t= - 2\kappa^2 \hat x_t dt 
                - 2\kappa d\hat{\cal W}^S_t,
\end{equation}
where $\kappa$ is a scalar constant.

{\it Proof}: 
Note that $S^{-1}=\Sigma S^\top \Sigma^\top$. 
Then, substituting $C=P^\top(-Z, I_n)$ for $B$ in 
Eq. \eqref{linear QSDE transfer}, we have 
\[
     B=\sqrt{2}
       \left( \begin{array}{cc} 
                - P_2 Y^{1/2} - P_1 Y^{-1/2}X  &  
                    P_1 Y^{-1/2}  \\
                -(YP_1+XP_2)Y^{1/2} - (XP_1-YP_2)Y^{-1/2}X~~~ &  
                    (XP_1-YP_2)Y^{-1/2}  \\
                 \end{array}\right), 
\]
where $P_1=\Re(P)$ and $P_2=\Im(P)$. 
First let us look at the (2,2) block matrix; 
since $X$ can take any symmetric matrix, here we set $X=0$, 
implying $YP_2Y^{-1/2}$ is independent on $Y$. 
This readily implies that $P_2$ has to be of the form 
$P_2=\sqrt{2}\kappa Y^{-1/2}$ with $\kappa$ a constant. 
Then, $XP_1Y^{-1/2}$ has to be independent on $X$ and $Y$. 
But as the (1,2) block matrix $\Theta:=P_1Y^{-1/2}$ also has 
to be independent on $X$ and $Y$, thus this is the case for 
$X\Theta$ as well. 
Then, $\Theta=0$ is only allowed, hence we obtain $P_1=0$. 
With these selection of $P_1$ and $P_2$, the (1,1) and (2,1) 
block matrices of $B$ take $-2\kappa I_n$ and zero, respectively. 
As a result, $B=-2\kappa I_{2n}$.

Next let us consider the matrix $A=\Sigma(G+\Im(C^\dagger C))$. 
From the above discussion, now we have 
$C=\sqrt{2}\kappa Y^{-1/2}(-Z, I_n)$, which leads to 
$A=\Sigma G - 2\kappa^2 I_{2n}$. 
This means that the matrix $G$ given in Eq. \eqref{G representation} 
must be independent on $X$ and $Y$. 
Then, similar to the above discussion, by setting $X=0$, we find 
that $G=(YRY, \Gamma Y^{-1}~;~Y^{-1}\Gamma^\top, R)$ has to be 
independent on $X$ and $Y$. 
But this requirement is only satisfied when $R=0$ and $\Gamma=0$; 
as a result, we have $G=0$. 
\hfill $\blacksquare$

This proposition states that, in order to dissipatively generate 
a desired pure Gaussian state in the state transfer setup, we 
are required to prepare identical and independent oscillators 
as memories. 
Note that any pure Gaussian cluster state (see the next section) 
can be effectively generated from the vacuum fields by applying 
suitably combined two-mode squeezing Hamiltonians and beam 
splitters \cite{Menicucci2007}, hence the proposed scheme is 
practical.


\section{Examples}

\subsection{Example 1: Gaussian cluster state generation}

It was shown in \cite{Menicucci2011} that the graph matrix 
$Z=X+iY$ can be used to capture several Gaussian graph states 
in a convenient graphical manner. 
In particular, the so-called canonical Gaussian cluster state 
\cite{Zhang2006,Menicucci2006,vanLoock2007,Menicucci2007}, 
which plays an essential role in continuous-variable one-way 
quantum computation, corresponds to 
\begin{equation}
\label{CV cluster}
   Z=X+ie^{-2\alpha}I_n, 
\end{equation}
where $X$ is the symmetric adjacency matrix representing the 
graph structure of the cluster state; 
for instance, the matrices 
\[
   X=\left( \begin{array}{cccc} 
         0 & 1 & 0 & 0 \\
         1 & 0 & 1 & 0 \\
         0 & 1 & 0 & 1 \\
         0 & 0 & 1 & 0 \\
      \end{array}\right),~~~
   X=\left( \begin{array}{cccc} 
         0 & 1 & 1 & 1 \\
         1 & 0 & 0 & 0 \\
         1 & 0 & 0 & 0 \\
         1 & 0 & 0 & 0 \\
      \end{array}\right),~~~
   X=\left( \begin{array}{cccc} 
         0 & 1 & 0 & 1 \\
         1 & 0 & 1 & 0 \\
         0 & 1 & 0 & 1 \\
         1 & 0 & 1 & 0 \\
      \end{array}\right)
\]
represent the chain, T-shape, and square structures, respectively 
(see Figure 2). 
On the other hand, $Y=e^{-2\alpha}I$ corresponds to the approximation 
error of the state with covariance matrix \eqref{pure CM general} 
to the ideal cluster state; 
that is, bigger $\alpha$ means that the state well approximates the 
ideal cluster state having the graph structure assigned by $X$.

\begin{figure}[t]
\centering
\begin{picture}(250,70)
\put(0,0)
{\includegraphics[width=3.5in]{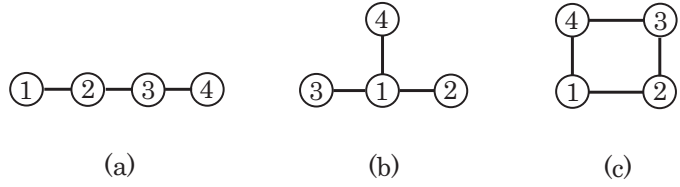}}
\end{picture}
\caption{
Typical cluster state with (a) chain, (b) T-shape, and (c) 
square structures. 
}
\end{figure}

Now, the nullifier dynamics \eqref{nullifier dynamics} is of 
the form 
\begin{equation}
\label{cluster nullifier dynamics}
    d\hat r_t=(Q^\top - e^{-2\alpha}P^\sharp P^\top)\hat r_t dt 
                 -2e^{-2\alpha}P^\sharp d\hat {\cal A}_t, 
\end{equation}
and the real part of the eigenvalue of the coefficient matrix 
$Q^\top - e^{-2\alpha}P^\sharp P^\top$ is $\Re(\lambda)
=-e^{-2\alpha}\|P^\top b\|$, 
where $b$ is the corresponding eigenvector. 
This means that making $\alpha$ bigger, or equivalently making 
the state more close to the ideal cluster state, renders 
the stability of the nullifier dynamics worse. 
Another observation from a more practical viewpoint is as follows; 
let us define the convergence time to the target by 
$T=1/{\rm min}|\Re(\lambda)|$ and denote the approximation error 
of the state to the ideal cluster one by $\epsilon=e^{-2\alpha}$. 
Then, it is straightforward to find $T\epsilon\geq c$ with $c$ 
a constant. 
Therefore, in order to dissipatively generate a pure Gaussian 
state that is very close to a desired cluster state, the 
convergence time has to be long.

More generally, as shown in \cite{Menicucci2011}, the matrices 
$X$ and $Y$ respectively correspond to the ideal and realistic 
parts of a Gaussian graph state in the sense that the covariance 
matrix of $(-X, I_n)\hat x$ is given by $Y/2$. 
That is, $Y$ can be regarded as the approximation error in 
approximating the ideal graph structure $X$. 
Therefore, the above-mentioned trade-off holds for a general 
Gaussian graph state. 
\hfill $\blacksquare$

\subsection{Example 2: Two-mode squeezed state}

There exist a number of proposals to generate a steady two-mode 
squeezed state in for instance atomic ensembles or 
nano-mechanical oscillators. 
The system matrices describing the two-mode squeezed state are 
given by $X=0$ and 
\[
    Y=\left( \begin{array}{cc}
           \cosh(2\alpha) & -\sinh(2\alpha) \\
           -\sinh(2\alpha) & \cosh(2\alpha) \\ 
      \end{array}\right), 
\]
where $\alpha$ denotes the squeezing parameter representing the 
degree of entanglement. 
In \cite{Cirac2,Polzik}, the dissipative channels achieving 
this goal was shown to be 
\begin{equation}
\label{channels for TMS}
    \hat L_1 = \mu \hat a_1 + \nu \hat a_2^*,~~~
    \hat L_2 = \mu \hat a_2 + \nu \hat a_1^*, 
\end{equation}
where $\mu=\cosh(\alpha)$ and $\nu=-\sinh(\alpha)$, while $\hat H=0$. 
In our formulation, this corresponds to setting \cite{KogaYamamoto}:
\[
    P=\left( \begin{array}{cc}
           i\cosh(\alpha) & i\sinh(\alpha) \\
           i\sinh(\alpha) & i\cosh(\alpha) \\ 
      \end{array}\right),~~~
    R=0,~~~
    \Gamma=0. 
\]
However, the dissipative channels \eqref{channels for TMS} are 
not easy to implement, since they are global and nontrivial 
coupling between the systems and the environment.

On the other hand, this dissipative system can be more easily 
implemented within the state transfer framework provided in 
Section 4, because we are only required to generate a two-mode 
squeezed state of optical fields and prepare two identical and 
independent oscillators. 
Note that a two-mode squeezed state of light can be effectively 
generated using a non-degenerate OPO. 
\hfill $\blacksquare$


\section{Conclusion}

In this paper, the dissipation-based state preparation method for 
general Gaussian case, which was originally formulated in 
\cite{KogaYamamoto}, was reconsidered in terms of the QSDE. 
This approach clarified that the nullifier dynamics of any Gaussian 
system generating a unique pure steady state is passive. 
As a byproduct, it was shown that there exists a trade-off between 
the closeness of the steady state to a given ideal graph state and 
the convergence time to that state. 
In addition, a convenient physical implementation method of 
a desired Gaussian dissipative system was provided; 
the scheme has the structure of quantum state transfer, which is 
a key ingredient in quantum information technologies.


\end{document}